\documentclass[%
 reprint,
superscriptaddress, 
 amsmath,amssymb,
 aps,
 pre
]{revtex4-1}

\usepackage[utf8]{inputenc}
\newcommand{\vect}[1]{\boldsymbol{#1}}

\usepackage[T1]{fontenc}
\usepackage{csquotes}
\usepackage{graphicx}
\usepackage{dcolumn}
\usepackage{bm}

\usepackage{float}

\usepackage{dcolumn}
\usepackage{bm}
\usepackage{mathptmx}
\usepackage{etoolbox}
\usepackage{relsize}

\usepackage[utf8]{inputenc}

\usepackage[T1]{fontenc}
\usepackage{csquotes}

\usepackage{graphicx}
\usepackage{dcolumn}
\usepackage{bm}

\usepackage{float}
\usepackage[dvipsnames]{xcolor}
\colorlet{green1}{green!40!black!100!}

\usepackage{dcolumn}
\usepackage{bm}
\usepackage{mathptmx}
\usepackage{etoolbox}
\usepackage{relsize}

\begin{document}

\preprint{ }

\title{
Adsorption of polyelectrolytes in the presence of varying dielectric discontinuity between solution and substrate
} 

\author{Hossein Vahid}
\affiliation{Department of Applied Physics, Aalto University, P.O. Box 15600, FI-00076 Aalto, Finland}
\affiliation{Department of Chemistry and Materials Science, Aalto University, P.O. Box 16100, FI-00076 Aalto, Finland}
\affiliation{Academy of Finland Center of Excellence in Life-Inspired Hybrid Materials (LIBER), Aalto University, P.O. Box 16100, FI-00076 Aalto, Finland}

\author{Alberto Scacchi}
\affiliation{Department of Applied Physics, Aalto University, P.O. Box 15600, FI-00076 Aalto, Finland}
\affiliation{Academy of Finland Center of Excellence in Life-Inspired Hybrid Materials (LIBER), Aalto University, P.O. Box 16100, FI-00076 Aalto, Finland}
\affiliation{Department of Bioproducts and Biosystems, Aalto University, P.O. Box 16100, FI-00076 Aalto, Finland}
\affiliation{Department of Mechanical and Materials Engineering, University of Turku, Vesilinnantie 5, FI-20014 Turku, Finland}

\author{Maria Sammalkorpi}
\affiliation{Department of Chemistry and Materials Science, Aalto University, P.O. Box 16100, FI-00076 Aalto, Finland}
\affiliation{Academy of Finland Center of Excellence in Life-Inspired Hybrid Materials (LIBER), Aalto University, P.O. Box 16100, FI-00076 Aalto, Finland}

\author{Tapio Ala-Nissila}
\email{tapio.ala-nissila@aalto.fi}
\affiliation{Department of Applied Physics, Aalto University, P.O. Box 15600, FI-00076 Aalto, Finland}
\affiliation{Quantum Technology Finland Center of Excellence, Department of Applied Physics, Aalto University, P.O. Box 15600, FI-00076 Aalto, Finland}
\affiliation{Interdisciplinary Centre for Mathematical Modelling and Department of Mathematical Sciences, Loughborough University, Loughborough, Leicestershire LE11 3TU, United Kingdom}

\begin{abstract}{
We examine the interactions between polyelectrolytes (PEs) and uncharged substrates at conditions corresponding to a dielectric discontinuity between the aqueous solution and the substrate. To this end, we vary the relevant system characteristics, in particular the substrate dielectric constant $\varepsilon_{\rm s}$ under different salt conditions. We employ coarse-grained molecular dynamics simulations with rodlike PEs in salt solutions with explicit ions and implicit water solvent with dielectric constant 
$\varepsilon_{\rm w} = 80$. As expected, at low salt concentrations, PEs are repelled from the substrates with $\varepsilon_{\rm s} < \varepsilon_{\rm w}$ but are attracted to substrates with a high dielectric constant due to image charges. This attraction considerably weakens for high salt and multivalent counterions due to enhanced screening. Further, for monovalent salt, screening enhances adsorption for weakly charged PEs, but weakens it for strongly charged ones. Multivalent counterions, on the other hand, have little effect on weakly charged PEs, but prevent adsorption of highly charged PEs, even at low salt concentrations. We also find that correlation-induced charge inversion of a PE is enhanced close to the low dielectric constant substrates, but suppressed when the dielectric constant is high. To explore the possibility of a PE monolayer formation, we examine the interaction of a pair of like-charged PEs aligned parallel to a high dielectric constant substrate with $\varepsilon_{\rm s} = 8000$. Our main conclusion is that monolayer formation is possible only for weakly charged PEs at high salt concentrations of both monovalent and multivalent counterions. Finally, we also consider the energetics of a PE approaching the substrate perpendicular to it, in analogy to polymer translocation. 
Our results highlight the complex interplay between electrostatic and steric interactions and contribute to a deeper understanding of PE-substrate interactions and adsorption at substrate interfaces with varying dielectric discontinuities from solution, ubiquitous in biointerfaces, PE coating applications, and designing adsorption setups.
}
\end{abstract}

\date{\today}

\maketitle

\section{introduction}

Polyelectrolytes (PEs), i.e., charged polymers, are of great interest in physics, materials science, chemistry, and biology.
PE adsorption onto substrates is a particularly important process with applications covering, e.g., water purification, production of paper and cardboard, energy storage~\cite{zhu2021}, chemical sensors~\cite{feng2010, kim2011}, stabilization of colloids~\cite{napper1983, eisenriegler1993}, and metal corrosion protection~\cite{andreeva2010}.
It has been the subject of extensive experiments~\cite{maroni2014, eneh2020, ivanov2021, li2023, hongwei2023}, theoretical investigations~\cite{netz1999, dirir2015, merlitz2015, xie2016, sahin2019}, and particle-based simulations~\cite{luque2014, de2015, kostritskii2016, sanchez2019, bagchi2020, wang2023}.
Biological examples of PEs are DNAs, RNAs, proteins, and microtubules.
Their interactions with bio-based systems inevitably lead to interactions of the former with the cellular membranes, which are composed of a chemically diverse set of lipids~\cite{langecker2014}.
Additionally, PE assemblies are essential from the point of view of biomedical applications, including vesicle transport~\cite{di2006, mclaughlin2005}, preventing bacterial adhesion~\cite{zhu2015, ouni2021}, and in drug and gene delivery systems~\cite{yin2014, maier2000, pott2002, pott2003, kahl2009, leal2008}. 

Interfaces are able to modify the structural and compositional characteristics of PEs in solutions as compared to their bulk state due to complex interplay between forces of different origins, such as steric effects and electrostatic interactions~\cite{jacob1992}. 
The relative permittivity of the substrate, $\varepsilon_{\rm s}$, can differ significantly from that of the solution, spanning a broad range of values~\cite{jacob1992}. If water at room temperature is considered as the solvent, the solvent relative permittivity $\varepsilon_{\rm r} = \varepsilon_{\rm w} \approx 80$. On the other hand, for common PE adsorption substrates such as mica, silica, and also carbon-based materials used for, e.g., DNA translocation, it varies between $2$ and $10$. Many oxides also commonly used as adsorption substrates exhibit relative permittivities between $15$ and $40$, while materials employed in supercapacitors can possess extremely high dielectric constants, such as calcium copper titanate reaching a relative permittivity of approximately $10 000$.

As PE solutions and PE assemblies are in contact with various substrates in many of the applications, a thorough understanding of the fundamental PE-substrate interactions is needed. These interactions are highly complex and depend on substrate polarization, solution characteristics, relevant electrostatic forces stemming from PE linear charge density, substrate surface charge density, co- and counterion valencies and image charges~\cite{bagchi2020}, steric and entropic effects (ion sizes, configurational entropies and PE flexibility~\cite{cherstvy2011, cherstvy2012, dias2012, chang2021}), all of which influence the overall adsorption behavior and stability of the system~\cite{elizarova2018, zhu2018}.
Straightforward electrostatic considerations show that for $\varepsilon_{\rm s} < \varepsilon_{\rm w}$, the image charges induced on the substrate are repulsive, while for $\varepsilon_{\rm s} > \varepsilon_{\rm w}$ the response is opposite, namely attractive. To overcome the repulsive image charges, many works on PE-substrate adsorption have focused on the case of adding attractive charges on the substrate~\cite{Dobrynin1997, netz1999, messina2004, cherstvy2011, cherstvy2012, bagchi2020, wang2023} or applying an external electric field~\cite{son2021}.
The situation for high-dielectric constant, metallic-character substrates is more subtle, however. Although the image charges are attractive, a competition between the interactions mentioned above exists~\cite{netz2002, cheng2004, seijo2009, lee2018}. While adsorption of isolated PEs on metallic-character substrates due to attractive images is possible, the spontaneous formation of dense layers is a nontrivial issue due to the complex interactions.

From the theoretical point of view, the simplest way to model PEs is based on the mean-field Poisson-Boltzmann (PB) approach.
However, large dielectric discontinuities and the presence of multivalent ions invalidate the PB approach due to enhanced charge correlations and fluctuations~\cite{blossey2018, hatlo2010, kanduvc2007, wang2015}. Mean-field theories often fall short of accurately representing many effects, including attraction between like-charged bodies~\cite{bowen1998, jho2008}, charge overcompensation~\cite{nguyen2000, besteman2004}, and finite sizes of ions~\cite{borukhov1997, quesada2003, li2009}. There have been attempts to modify the PB theory by incorporating ion-size effects~\cite{heyda2012,zhou2011, colla2017, batys2017, vahid2022, yang2022}.
However, to effectively address the impact of image charges and presence of multivalent ions requires going beyond the mean-field approximation, e.g., adopting a theoretical framework capable of managing strong-coupling electrostatic interactions~\cite{buyukdagli2016, buyukdagli2014, jho2008}. Alternatively, classical density functional theories~\cite{gonzalez2018, cats2021, cats2022, bultmann2022} or particle-based simulation techniques, such as Monte Carlo~\cite{javidpour2019, guerrero2014} or molecular dynamics (MD)~\cite{yuan2020} simulations can be employed.
Extensive studies have explored the adsorption of PEs to charged substrates, yet the effect of image charges is often overlooked~\cite{wang2010, de2016, caetano2017, chang2022, yuan2024}.
However, recent theoretical and computer simulation studies have revealed nontrivial effects of substrate polarizability in the interfacial behavior of both ions and PEs~\cite{bagchi2020, son2021, levin2009, wang2023}.
In the absence of surface charges, the impact of image charges is significant.

In this paper, we employ coarse-grained MD simulations to study PE solutions at electrostatically uncharged planar interfaces. 
We consider a simplified model of a rigid
double-stranded B-DNA-like molecule~\cite{cherstvy2005} that experiences a dielectric contrast between the solution and an impenetrable substrate. 
We scrutinize the interactions and adsorption processes of a single and a pair of identical rodlike PEs with equal charge and address questions that arise regarding how substrate polarization can influence the adsorption phenomena, which are pronounced in the presence of multivalent ions in the salt. 
We focus on the effect of salt concentration, ion valency, and substrate characteristics, which are key aspects controlling adsorption as observed in experimental settings.
For example, in the case of PE multilayers, one of the most influential variables on film thickness is the concentration of salt in the deposition solution~\cite{decher1992, dubas1999}.
We also anticipate that correlation effects will be significant, particularly in cases involving asymmetric salts that feature a low concentration of an ionic species with high valency~\cite{Antila2017}. The experimental relevance of this has been noted in previous studies~\cite{rau1992, rau1992-2}.

The outline of the paper is as follows.
In Sec.~\ref{sim_model}, we present an overview of the system setup, the polymer, substrate, and electrolyte model, as well as the computational details. 
In Sec.~\ref{sec:results_discussion}, we present our key findings. In the first place, we show the influence of surface polarization in systems comprised solely of ions. 
Next, we introduce one and then two PEs into the system, forming the core part of this section. Finally, this section also discusses the interactions of PEs oriented perpendicular to the substrate, a configuration of experimental relevance for, e.g., DNA translocation~\cite{palyulin2014, buyukdagli2019}. 
Finally, in Sec.~\ref{conclusion} we conclude by summarizing our findings.

\section{Simulation model}\label{sim_model}

\begin{figure}[t!]
\centering
  \includegraphics[width=1\linewidth]{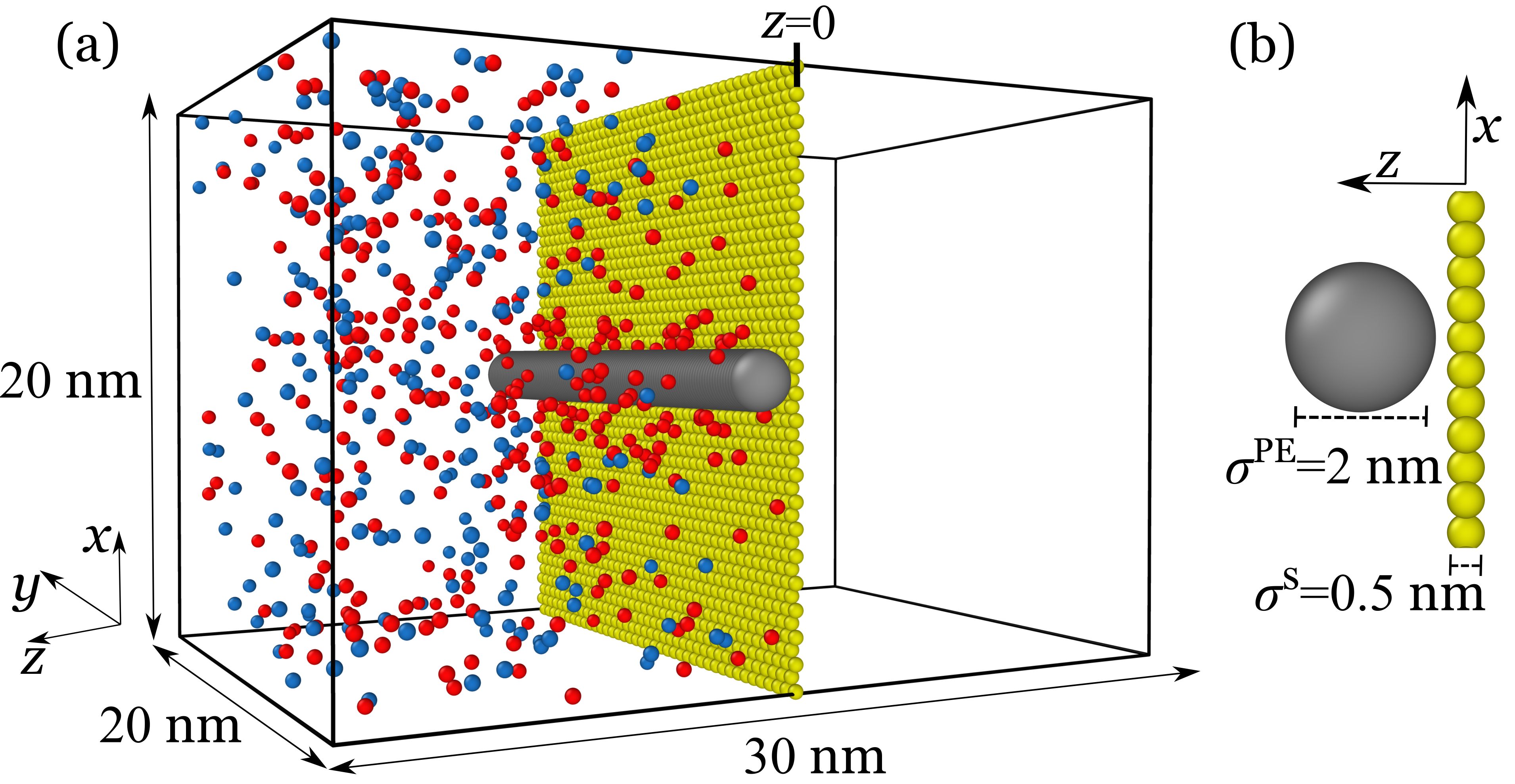}
 \caption{(a) A snapshot of the simulation box, measuring $20 \times 20 \times 30$~nm\(^3\), with periodic boundary conditions along the \(x\) and the \(y\) axes. An impenetrable substrate, depicted in yellow with dielectric constant of \(\varepsilon_{\rm s}\), is positioned at \(z=0\) where the force centers of the beads are. The substrate fills the whole negative $z$ side. A hard wall is located at \(z=15\)~nm, resulting in the ions bouncing back and not leaving the top half of the simulation box. The box contains a PE (gray), monovalent counterions and cations (red), and monovalent coions (blue) at $Z^{\rm c}c^{\rm c}=60$ mM here. (b) Close-up view of the PE and the bead-based substrate with the PE diameter of $\sigma^{\rm PE} = 2$ nm and surface bead diameter of $\sigma^{\rm s} = 0.5$ nm shown. The substrate is oriented parallel to the $xy$ plane and structured in a triangular lattice.}
 \label{Fig-schematic}
\end{figure}

Our simulation setup contains a negatively charged rodlike PE and explicit ions immersed in a medium of dielectric constant $\varepsilon_{\rm w} = 80$, corresponding to implicit water as the solvent. The simulation box is rectangular with fixed dimensions $L_{x}= L_{y}=20$ nm. The length of the inaccessible region of the system is fixed along the $z$ axis at 15 nm. On the other hand, depending on the orientation of the PE, namely parallel or perpendicular to the substrate, the accessible region along the $z$ axis varies between 15 and 20 nm, respectively. The substrate is fixed at $z=0$, confining the PE and the free ions to the $z>0$ side of the system (cf. Fig.~\ref{Fig-schematic}).

The PE chain consists of spherical beads, each with a diameter $\sigma^{\rm PE}=2$ nm and carrying a charge of $-e$. The center-to-center separation between consecutive beads is $0.17$ nm, resulting in a linear charge density $\lambda_0=-6$ $e/$nm, close to that of DNA molecules~\cite{kominami2019}.
When the PE is aligned parallel to the substrate, it spans a length of $20$ nm and comprises $120$ beads, whereas in the perpendicular alignment, it measures $10$ nm in length and is made up of $60$ beads.
To neutralize the charge of the PE, we add monovalent counterions. The system response is examined for added salt concentrations ranging from $30$ mM to $500$ mM. 
We systematically alter the valency of added cations ($Z^{\rm c}$) and anions ($Z^{\rm a}$) to thoroughly investigate the system response to varying ion correlations. In each system, we ensure charge neutrality by maintaining the condition $Z^{\rm c}c^{\rm c}=\lvert Z^{\rm a}\lvert c^{\rm a}$, where $c^{\rm c}$ and $c^{\rm a}$ represent the concentrations of cations and anions, respectively.

The impenetrable substrate, parallel to the $xy$ plane, is composed of spherical beads arranged in a triangular lattice, each with a diameter of $\sigma^{\rm s}=0.5$ nm. The substrate implicitly extends in the whole $z<0$ region. The relative permittivity of the substrate is varied to mimic different types of materials, namely $\varepsilon_{\rm s} = 2$ corresponding to an insulating material, $80$ to mimic a nonpolarizable substrate, and higher values, such as $500$, $1000$, and $8000$ to represent conductive substrates.

Due to the dielectric mismatch between the substrate and water, electric charges are induced on the substrate plane at $z=0$ in response to the presence of ions and PEs within the implicit water environment.
The polarization charge density at each point on the substrate in response to the electric field at that point $\vect{E^i}$, generated by the PE solution and induced charges on other points on the substrate surface, is given by~\cite{jackson1999}
\begin{equation}\label{eq:induced}
    \sigma_{\rm p}^i = 2\varepsilon_0\Delta{\varepsilon}~\vect{n}\cdot\vect{E^i}, \hspace{30pt}  i\in {\rm s},
\end{equation}
where $\Delta{\varepsilon}=({{\varepsilon_{\rm s}-\varepsilon_{\rm w}}})/({{\varepsilon_{\rm s}+\varepsilon_{\rm w}}})$ is the dielectric mismatch between the substrate and the implicit solvent (water), $\varepsilon_{\rm 0}$ is the vacuum permittivity, and $\vect{n}$ is the substrate normal vector toward the solvent.
Here, the parameter $\Delta{\varepsilon}$ has values of $-0.95$, $0$, $0.72$, $0.85$, and $0.98$ for $\varepsilon_{\rm s}$ equal to $2$, $80$, $500$, $1000$, and $8000$, respectively.

The interactions among the PE beads, substrate beads, and ions are governed by the Weeks-Chandlers-Andersen potential~\cite{anderson}, given by
 
\begin{equation}
    U^{ij}(r)=4\epsilon\left[\left(\frac{\sigma^{ij}}{r}\right)^{12}-\left(\frac{\sigma^{ij}}{r}\right)^{6}+\frac{1}{4}\right]H(r_{\rm c}^{ij}-r),
\end{equation}
where $r$ represents the distance between two interacting particles $i$ and $j$, and $H(r)$ is the Heaviside function, truncating the potential at $r_{\rm c}^{ij}=2^{1/6}\sigma^{ij}$. The depth of the potential well is set to $\epsilon=0.1$ kcal mol$^{-1}$.
Here, the interspecies diameter, $\sigma^{ij}$, for the particle pair $i, j$ is calculated as the arithmetic mean of their diameters, defined as $\sigma^{ij} = (\sigma^i + \sigma^j)/2$.
Both cations and anions have the same hydrated diameter of $\sigma^{\rm c}=\sigma^{\rm a}=0.5$ nm.

Additionally, we calculate electrostatic interactions in real space with a cutoff of $15$ nm (extending to 20 nm when the PE is perpendicular to the substrate) using a pairwise Coulombic potential, defined as 
\begin{equation}
U_{\rm C}^{ij}(r)=\frac{ Z^{i} Z^{ j} \ell_{\rm B}}{\beta e r}.    
\end{equation}
Here, $Z$ represents the charge valency. The notation $Z^{\rm c}:\vert Z^{\rm a}\vert$ is used to represent the valencies of cations and anions, respectively. The Bjerrum length in water at $300$ K is $\ell_{\rm B}=\beta e^2/(4\pi \varepsilon_0\varepsilon_{\rm w})\approx 0.7$ nm, where $\beta = 1/k_{\rm B}T$, with $k_{\rm B}$ being the Boltzmann constant and $T$ the temperature of the system.
The Debye screening length, defined as $\kappa^{-1}=(4\pi\ell_{\rm B}\Sigma_{i}(Z^{i})^2c^{i})^{-1/2}$, represents the distance over which electrostatic interactions are exponentially screened by salt ions.
For long-range electrostatic interactions, we employ the Particle-Particle Particle-Mesh (P$^3$M) summation procedure~\cite{plimpton1997} with slab corrections in the $z$ direction and a relative force accuracy of $10^{-5}$. 
$z$-periodicity is applied by adding empty space between the repetitions of the simulation box~\cite{yeh1999}. The proportion of the extended dimension in the $z$ direction relative to the original dimension is set at $3$.
A reflective wall is positioned at $z = 15$ nm (or at $z=20$ nm for the PE perpendicular to the substrate). Periodic boundary conditions are applied in the $x$ and $y$ directions.
All simulations are performed using LAMMPS {\it 2 Aug 2023} stable release~\cite{plimpton1995, brown2009, Thomson2022} in the canonical ($NVT$) ensemble, employing the Nos\'e-Hoover~\cite{nose1984, hoover1985} thermostat to maintain $T = 300$ K. The integration time step is set to $2$ fs, and the trajectories are sampled every $2$ ps over a $20$ ns simulation period, excluding the initial $5$ ns for equilibration.
The initial configurations of the simulations are generated using Moltemplate~\cite{jewett2021} and Packmol~\cite{packmol}. 

We also incorporate polarizability due to the dielectric mismatch using the ICC$^*$ (Induced Charge Computation) algorithm~\cite{tyagi2010}, which has been implemented into LAMMPS as detailed in Ref.~\onlinecite{nguyen2019}.
In the algorithm, the substrate surface ${\rm s}$ is discretized into surface elements.
A charged particle $i$ interacts with ${\rm s}$ via the surface element charges that capture the effect of the image charges due to the particle itself and all other charges in the system, including all the ions and the PE charges. The resulting coupled equations are solved self-consistently.

To quantify the substrate-PE and PE-PE interactions, we calculate the normalized mean force as a function of their separation distance, $l$, which is defined as $f(l) = F(l) / L$, with $L$ representing the length of the PE. The term $F(l)$ refers to the average total force observed either between the substrate and PE or between two PEs over the production runs for different specific distances $l$. The corresponding potential of mean force $V$ can be calculated via $V(l) = \int_l^\infty F(l')\,dl'$.

\section{Results and discussion}\label{sec:results_discussion}
\subsection{Single PE parallel to the substrate}
To check the effect of image charges on the ion distributions, we first performed simulations for systems consisting exclusively of solvated ions without the presence of any PEs.
Figure~S1 of the Supplementary Material (SM) displays the number density profiles of cations and anions in a $1:1$ salt solution with a concentration of $250$ mM.
We observe that near the high dielectric constant substrate corresponding to a metallic character, both cations and anions have higher densities as compared to those close to low dielectric discontinuity substrate, which does not exhibit significant polarization. Conversely, near a low dielectric constant substrate (with insulating character), the ion densities are lower than those observed near a non-polarizable substrate~\cite{vertenstein1987, frydel2011}.
This phenomenon can be attributed to the interaction of ions with their image charges. When ions are close to a substrate bearing metallic characteristics, they encounter image charges of opposite signs, leading to attraction. In contrast, for a low dielectric constant, a relatively insulating substrate, the image charges mirror the sign of the ions, resulting in repulsion.  

\subsubsection{Influence of added monovalent salt}

\begin{figure}[t!]
\centering
  \includegraphics[width=1\linewidth]{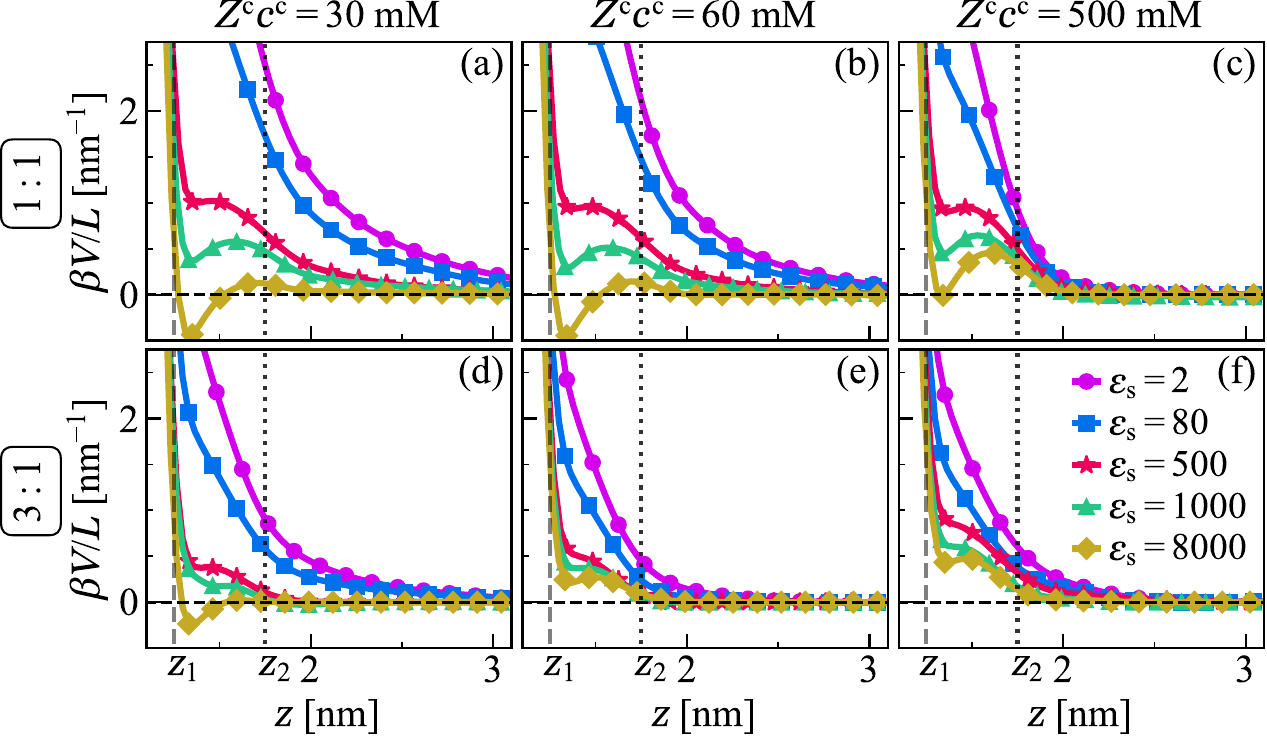}
 \caption{The normalized potential of mean force $\beta V/L$ as a function of the distance $z$ from the center of the PE to the substrate plane at $z=0$ for salts with valency ratios of (a)-(c) $1:1$ and (d)-(f) $3:1$, while varying both the salt concentration $Z^{\rm c}c^{\rm c}$ and the relative dielectric constant of the substrate $\varepsilon_{\rm s}$ (see Fig.~S2 of SM for $2:1$ salt). The vertical dashed lines represent the distance at which the surface of the PE is in contact with the substrate beads $z_1=(\sigma^{\rm PE}+\sigma^{\rm s})/2$, while the dotted lines represent the distance at which they are one ion diameter apart such that $z_2=z_1+\sigma^i$.}
 \label{Fig1}
\end{figure}

Next, we examine the influence of added salt on the interactions between a DNA-like PE and a substrate. 
Figure~\ref{Fig1} illustrates the potential of mean force $V$ on the PE as a function of the distance between the PE and the substrate while varying both the salt concentration $Z^{\rm c}c^{\rm c}$ and the dielectric constant $\varepsilon_{\rm s}$.
When $\varepsilon_{\rm s}<\varepsilon_{\rm w}$, induced charges, resulting from the presence of the PE, bear the same charge as the PE.
Consequently, the PE is repelled from the substrate, a phenomenon observed with DNA and neutral phospholipid membranes (typically having a relative dielectric permittivity of $2-4$), both experimentally~\cite{gromelski2006} and in MD simulations~\cite{antipina2016}. This implies that in PE adsorption onto low dielectric constant substrates, the substrate must carry static charges~\cite{cherstvy2012}.
Theoretical studies based on a one-loop expansion of the self-consistent equation and incorporating charge correlations have revealed that strongly charged PEs are attracted to highly like-charged substrates due to the charge correlations~\cite{buyukdagli2016-3, sahin2019}. However, increasing the bulk salt concentration leads to the unbinding of the PE from the substrate~\cite{buyukdagli2016-3}.
Weakly charged PEs are repelled by like-charged substrates with $\varepsilon_{\rm s}<\varepsilon_{\rm w}$, irrespective of salt concentration and substrate static charge density~\cite{buyukdagli2016-3}.
Similarly, mean-field-based investigations have shown that strongly charged semiflexible PEs desorb from oppositely charged planar substrates with $\varepsilon_{\rm s}<\varepsilon_{\rm w}$ at low and high monovalent salt concentrations due to image charge and entropic effects, respectively~\cite{netz1999}.
On weakly charged substrates, repulsive interactions originating from image charge effects keep strongly charged PEs away from insulating character substrates~\cite{buyukdagli2016-3}.
In contrast, when $\varepsilon_{\rm s}>\varepsilon_{\rm w}$, the induced charges change sign, and the PE is attracted to the substrate.
In all cases, increasing salt concentration results in more effective screening of the image charge effects, thus reducing the interaction strength between the PE and the substrate.
This is consistent with the experimental findings reported in Ref.~\onlinecite{boulmedais2006}, which demonstrated that increasing the salt concentration leads to a reduction in the adsorption of poly(L-lysine) and heparin on conductive indium tin oxide substrate.
It has been shown, by means of Debye-H{\"u}ckel theory, that at high concentrations of monovalent salt, semiflexible PEs partially or fully detach from oppositely charged spheres~\cite{netz1999-2, boroudjerdi2005} and planar substrates~\cite{netz1999} due to the screening of electrostatic interactions. At approximately $1$~M salt concentration, the PE and the sphere fully separate and reach equilibrium~\cite{netz1999-2}. However, PEs form a lamellar structure on the substrate at intermediate salt concentrations, leading to substrate overcharging~\cite{netz1999, netz1999-2}.

Additionally, when the gap between the PE and the substrate surpasses the ion diameter ($r>1.75$ nm), $V$ converges to zero at all salt concentrations.
When the PE is almost touching the substrate (i.e., $r<(\sigma^{\rm s}+\sigma^{\rm PE})/2=1.25$ nm), strong repulsion emerges due to the steric repulsion.
At a separation approximating one ion diameter ($r\approx\sigma^{i}$), the mean force potential impacting the PE near the relatively metallic substrates with $\varepsilon_{\rm s}=8000$ shows an energy barrier. This can be explained by noting that at such a distance, the ions condensed to the PE side facing the substrate establish a steric barrier that influences the interactions between the PE and the substrate.
In addition, the adsorption of ions in this region reduces the positive charges on the substrate originally induced due to the presence of PE, which in turn weakens the attractive forces acting on the PE.
As the salt concentration increases, the height of this barrier also increases. This suggests that in systems with a substrate with metallic character, a higher ion concentration between the substrate and the PE at these proximities enhances the steric barrier effect. However, as noted earlier, the increased salt concentration simultaneously leads to a diminished attraction between the PE and the substrate. 

\subsubsection{Influence of multivalent counterions}

We next focus on the influence of multivalent ions on PE-substrate interactions. Incorporating multivalent cations into the electrolyte significantly reduces PE-substrate electrostatic interactions. This implies that higher-charged cations are more efficient in screening electrostatic interactions between the PE and the substrate. 
For example, for $\varepsilon_{\rm s}=8000$, the introduction of trivalent cations results in a weaker attraction between the PE and the substrate than that observed with monovalent cations, as evidenced by comparing Figs.~\ref{Fig1}(a) and \ref{Fig1}(i).
Our findings are consistent with the experimental observations for DNA-cationic substrates, where adding multivalent cations releases DNA from the substrate~\cite{koltover2000}.
Additionally, in $1:\vert Z^{\rm a}\vert$ salts, we note that an increase in anion valency does not substantially alter the interactions between the PE and the substrate, as demonstrated in Fig.~S3 of the SM.

On the other hand, one might expect to have like-charge attraction between the PE and the substrate with $\varepsilon_{\rm s}<\varepsilon_{\rm w}$ in solutions with multivalent ions, see e.g. Refs.~\onlinecite{grosberg2002, moreira2000, buyukdagli2017, vahid2023} for detailed discussion.
Theoretical studies, such as those using a one-loop expansion of self-consistent theory, have demonstrated that highly charged PEs experience stronger adsorption onto insulating substrates with like static charges in the intermediate coupling regime~\cite{sahin2019}.
Furthermore, Ref.~\onlinecite{buyukdagli2019-2} introduced a one-loop-dressed strong coupling theory to examine PE adsorption onto like-charged substrates in solutions containing tri- and tetravalent counterions with monovalent salt. The findings revealed that increasing counterion valency can enhance the adsorption of PEs, even on weakly charged substrates.
Note that in these studies, the substrates have fixed charges that remain unaffected by the presence of ions and PEs. However, Fig.~\ref{Fig1} shows that the PE attraction to the substrate with $\varepsilon_{\rm s}=8000$ is weakened by introducing multivalent cations to the salt. In our simulations, the bare charge density of the substrate is always zero, and the introduction of multivalent ions reduces, and in some cases completely neutralizes, the induced charges on the substrate. Consequently, the PE-substrate attraction significantly decreases. At high enough concentrations ($Z^{\rm c}c^{\rm c}\geq60$ mM), the attraction completely vanishes.

Moreover, in cases where $\varepsilon_{\rm s}=80$, the absence of induced charges in the system does not preclude the PE from experiencing a repulsive force when in proximity to the substrate. This repulsion is primarily attributed to the steric forces arising from ions condensed on the side of the PE that faces the substrate.
As the PE moves closer to it, closer than the LJ interaction cutoff distance, the magnitude of this repulsive force is observed to intensify. 
The repulsion is barely diminished as the salt concentration rises due to the attractive depletion forces generated by the osmotic pressure gradient between the bulk and the gap region separating the PE and the substrate. Yet this attractive force cannot overcome the steric and LJ repulsive forces.

\begin{figure}[t!]
\centering
  \includegraphics[width=1\linewidth]{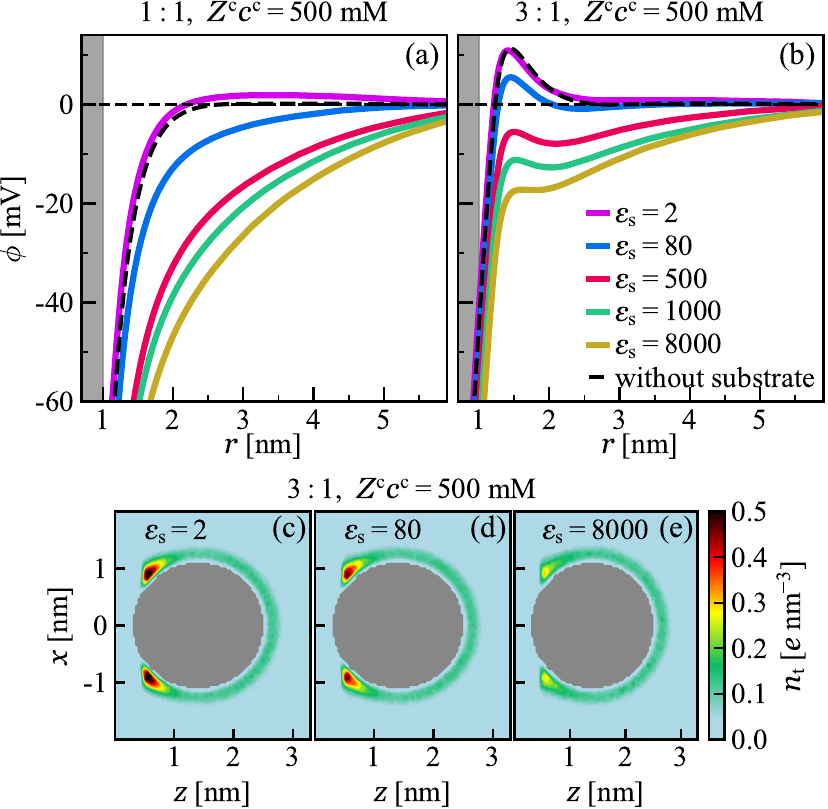}
 \caption{The average electrostatic potential $\phi$ (cf. Eq.~(5)) as a function of radial distance $r$ from the PE center in (a) $1:1$ and (b) $3:1$ salt solutions with $Z^{\rm c}c^{\rm c}=500$ mM. Here, the PE is at $z=1.4$ nm. The gray bar indicates the PE with a radius of $1$ nm. (c)-(e) The distribution of the total two-dimensional charge density, $n_{\rm t}=n^{\rm c}+n^{\rm a}$, in $3:1$ salt at $Z^{\rm c}c^{\rm c}=500$ mM for different $\varepsilon_{\rm s}$. The data are averaged along the $y$ axis. The center of the PE axis is at $x=0$ and $z=1.4$ nm, with alignment along the $y$ axis. The PEs are shown in gray.}
 \label{cumulative_charge}
\end{figure}

The results presented in Fig.~\ref{Fig1}(f) indicate that the potential curves for both high dielectric constant, metallic-character substrates, and those that have low dielectric constants tend to converge towards those of a non-polarizable substrate ($\varepsilon_{\rm s}=80$) with highly multivalent cations and at the highest salt concentrations studied here. This further implies that the screening of electrostatic interactions between the substrate and the PE is significantly more effective with high salt concentrations and cation valencies.
In all cases, $V$ is closer to those with $\varepsilon_{\rm s}=80$ when $\varepsilon_{\rm s}<\varepsilon_{\rm w}$ than when $\varepsilon_{\rm s}>\varepsilon_{\rm w}$. 
This indicates that substrates with a low dielectric constant have a less pronounced effect on altering interactions between the PE and the substrate as compared to high dielectric constant substrates that have metallic character in their response.
Additionally, Fig.~\ref{Fig1}(f) illustrates that the potential of mean force for $\varepsilon_{\rm s}=8000$ approaches zero at separations $z \approx 1.5$ nm, which corresponds to less than one ion diameter from the substrate. Here, dense packing of multivalent counterions around the PE results in induced charges on the substrate that effectively neutralize those caused by the PE proximity.
Additionally, a possible explanation for the lower steric barrier height in $3:1$ salt as compared to $1:1$ salt across all values of $Z^{\rm c}c^{\rm c}$ when $\varepsilon_{\rm s}=8000$ could be attributed to the significantly lower number of ions present in the high-valency salt.

In Fig.~\ref{cumulative_charge}, we present the profiles of the mean electrostatic potential $\phi$ at $Z^{\rm c}:1$ salt when the PE is at $z=1.4$ nm from the substrate plane. This value corresponds approximately to the position of the global minimum in the potentials of Fig.~\ref{Fig1} (where such minima exist). 
In the case of $1:1$ salt, charge reversal and the associated reversal of $\phi$ are absent for the PE near high dielectric constant or non-polarizable substrates. However, a weak inversion of $\phi$ is observed when the PE is close to a low dielectric constant, insulating substrate with a dielectric constant of $\varepsilon_{\rm s}=2$, as the induced charges on the substrate, which have the same sign as the PE, also attract cations. 

The key to successful deposition of PE multilayers in a layer-by-layer fashion here would be charge inversion that should occur after each adsorption step. 
In $3:1$ salt and absence of any substrates (dashed line in Fig.~\ref{cumulative_charge}(b)), the overcompensation of DNA charge by trivalent cations condensed on the PE leads to the charge reversal of the PE and, consequently, the reversal of $\phi$~\cite{yang2023, vahid2024}.
However, the proximity of the substrate impedes full condensation of cations around the PE. This spatial limitation results in a diminished potential when the PE is near a non-polarizable substrate, as compared to the case where no substrate is present (dashed line).
In addition, we observe a notable effect of different substrate types on the charge reversal of the PE. 
the potential undergoes rapid decrease as a function of $\varepsilon_{\rm s}$. 
The proximity of low dielectric constant substrates to the PE tends to enhance charge reversal, whereas high dielectric constant, metallic-character substrates appear to suppress it. This phenomenon can be explained by considering the nature of induced charges on the low dielectric constant substrates. These charges have the same sign as the PE, attracting more counterions to the vicinity of the substrate and facilitating charge reversal. In contrast, the induced charges on high dielectric constant, metallic-character substrates, being opposite in sign to those of the PE, repel nearby counterions, thereby suppressing charge reversal.
This can be seen in the total ion charge profiles presented in Figs.~\ref{cumulative_charge}(c)-(e). The density of positive charges around the PE close to the substrate with $\varepsilon_{\rm s}=2$ is higher than that with $\varepsilon_{\rm s}=8000$, which leads to charge reversal, and consequently the reversal of the electrostatic potential.  
Additionally, we have calculated the value of $\phi$ in cases when enough space exists for a layer of counterions between the PE and the substrate (see Fig.~S4 of SM). 
We observe that under these conditions, the results for different values of $\varepsilon_{\rm s}$ converge to those observed for non-polarizable substrates.

\subsubsection{Influence of PE linear charge density}

\begin{figure}[t!]
\centering
  \includegraphics[width=1\linewidth]{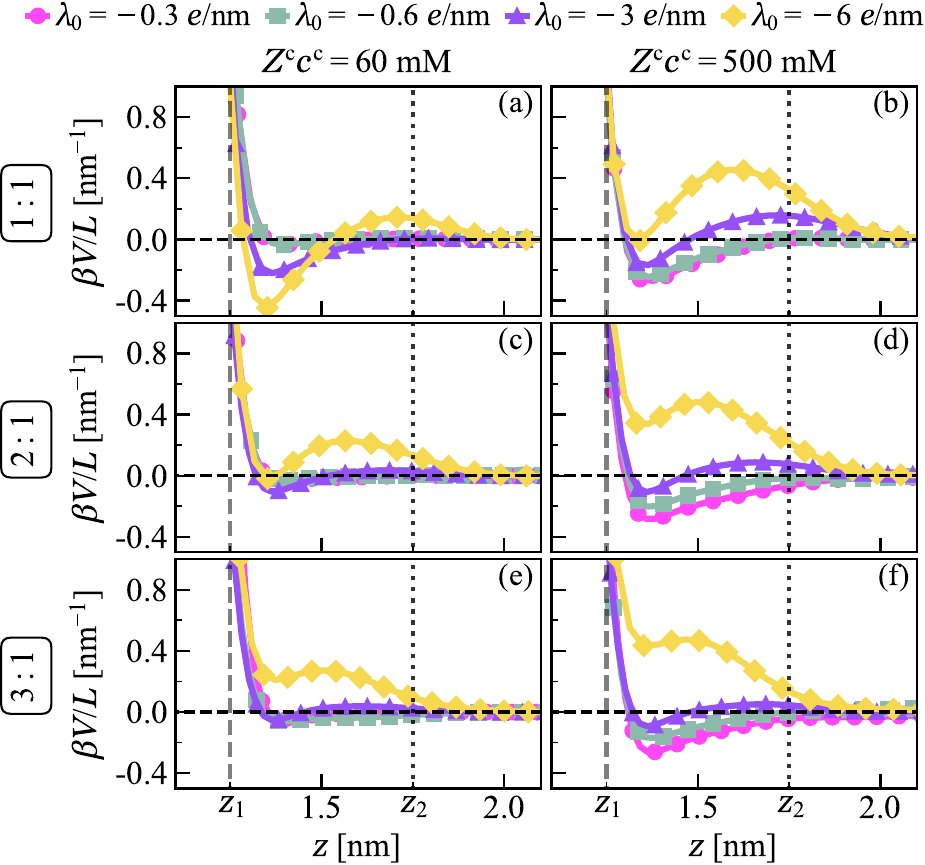}
 \caption{Normalized potential of mean force ($\beta V/L$) versus distance between the center of the PE and the substrate plane in (a)-(b) $1:1$, (c)-(d) $2:1$, (e)-(f) and $3:1$ salt solutions, across varying salt concentrations $Z^{\rm c}c^{\rm c}$. Here, $\varepsilon_{\rm s}=8000$. The vertical dashed lines represent the distance at which the surface of the PE is in contact with the substrate beads $z_1=(\sigma^{\rm PE}+\sigma^{\rm s})/2$, while the dotted lines represent the distance at which they are one ion diameter apart such that $z_2=z_1+\sigma^i$.}
 \label{lambda_PE}
\end{figure}

Figure~\ref{lambda_PE} represents the potential of mean force $\beta V/L$ between the PE and the high dielectric constant, metallic-character substrate with $\varepsilon_{\rm s}=8000$ as a function of the distance between the substrate plane and the center of the PE while varying the PE line charge, $\lambda_0$, and salt concentration $Z^{\rm c}c^{\rm c}$.
The first observation is that for weakly charged PEs, there is no energy barrier upon approaching the substrate. This can be attributed to the lower ion attraction to a weakly charged PE, resulting in fewer ions on the side of the PE facing the substrate. Consequently, this leads to a less pronounced steric barrier, which becomes large for strongly charged PEs.
The results presented in panel (a) also indicate that at low monovalent salt concentrations, the attractive minimum of $V$ strengthens with increasing $\lambda_0$. Interestingly, the trend reverses at high salt concentrations (panel (b)), showing a large barrier for highly charged PEs and an attractive minimum for weakly charged ones.
Thus, PEs in monovalent salt fall into two distinct categories, namely screening-enhanced adsorption for weakly charged PEs and screening-reduced adsorption for strongly charged PEs.

As already discussed regarding strongly charged PEs ($\lambda_0=-6$ $e/$nm) in Fig.~\ref{Fig1}, the presence of multivalent cations in salt leads to more effective screening of induced charges as well as the PE charge, thereby weakening the electrostatic attraction towards the substrate. In the case of $3:1$ salt, the highly charged PE is completely repelled from the substrate.
Interestingly, multivalent cations do not significantly influence the interaction of weakly charged PEs ($\lambda_0=-0.3$ $e/$nm) with the substrate, and adsorption is possible with higher salt concentrations.

\subsection{Two PEs parallel to a high dielectric constant substrate}
\begin{figure}[b!]
\centering
  \includegraphics[width=1\linewidth]{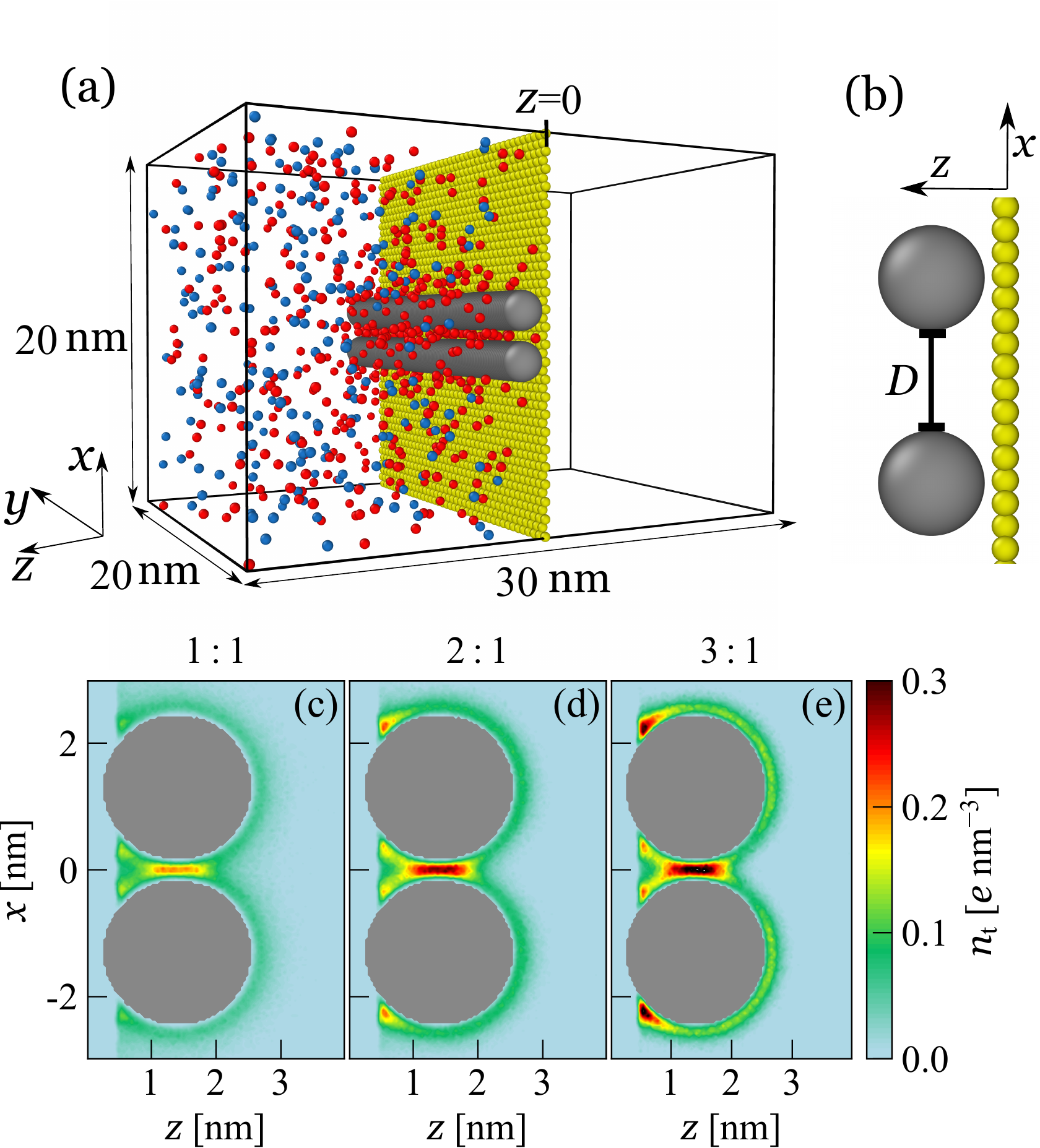}
 \caption{(a) A snapshot of the simulation box with two parallel identical PEs whose surfaces are separated by $D$ (their center-to-center separation is $D+\sigma^{\rm PE}$), measuring \(20 \times 20 \times 30\) nm\(^3\) with periodic boundary conditions along the \(x\) and \(y\) axes. The substrate, depicted in yellow and with the dielectric constant of \(\varepsilon_{\rm s}=8000\), is positioned at \(z=0\), while a hard reflecting wall is located at \(z=15\) nm. The system also contains monovalent counterions (red) and monovalent coions (blue) at $Z^{\rm c}c^{\rm c}=\vert Z^{\rm a}\vert c^{\rm a}=60$ mM. (b) Close-up view of the PEs oriented along the $y$ axis and the bead-based substrate which is along the $xy$ plane at $z=0$ and structured in a triangular lattice. $D$ is the PE-surface-to-PE-surface distance. (c)-(e) Panels depicting the total two-dimensional charge distribution $n_{\rm t}=n^{\rm c}+n^{\rm a}$ for cations and anions in $Z^{\rm c}:1$ salt at $Z^{\rm c}c^{\rm c}=\vert Z^{\rm a}\vert c^{\rm a}=500$ mM. The data are averaged along the $y$ axis. The PEs are at $z=1.4$ nm along the $y$ axis, with an equilibrium spacing of $D=0.6$ nm between them. The PEs are shown in gray. }
 \label{Fig2-schematic}
\end{figure}

We next explore systems comprising two identical PEs parallel to a high dielectric constant, metallic-character substrate. These PEs are positioned at their equilibrium distance from the substrate ($z=1.4$ nm, obtained for single PE interacting with the substrate, see Fig.~\ref{lambda_PE}) with $\varepsilon_{\rm s}=8000$. The two PEs have identical linear charge density, either $\lambda_0=-6$ $e/$nm or $-0.3$ {\it e}/nm, and their surfaces are separated from each other by distance $D$ along the $x$ axis (see Fig.~\ref{Fig2-schematic}(a)).
In Figs.~\ref{Fig2-schematic}(c)-(e), we show the total ion charge profiles. As expected, higher cation valency results in an increased positive charge condensation between the two PEs. This trend, consistent with findings in other studies without a substrate~\cite{grosberg2002, moreira2000, buyukdagli2017, vahid2023}, indicates that enhanced counterion correlation contributes to the attraction between identical PEs.

\subsubsection{PE-PE interactions on a high dielectric constant substrate}

To examine the influence of the substrate on PE-PE interactions, we fix the distance of the PEs from the substrate. The potential of mean force $V$ between the PEs is then calculated as a function of their surface-to-surface separation $D$.
Figure~S6 in the SM demonstrates that the equilibrium distance between PEs remains constant regardless of their distance to the substrate, while a slight decrease of approximately $0.02\ k_{\rm B}T/$nm is observed in their attraction strength as the PEs approach the substrate.
These findings suggest that to determine the equilibrium distance of a two-PE system from the substrate, we can initially position the PEs at their equilibrium separation from each other, $D=0.25$ nm for $\lambda_0=-0.3$ $e/$nm and $D=0.55$ nm for $\lambda_0=-6$ $e/$nm.

\begin{figure}[b!]
\centering
  \includegraphics[width=1\linewidth]{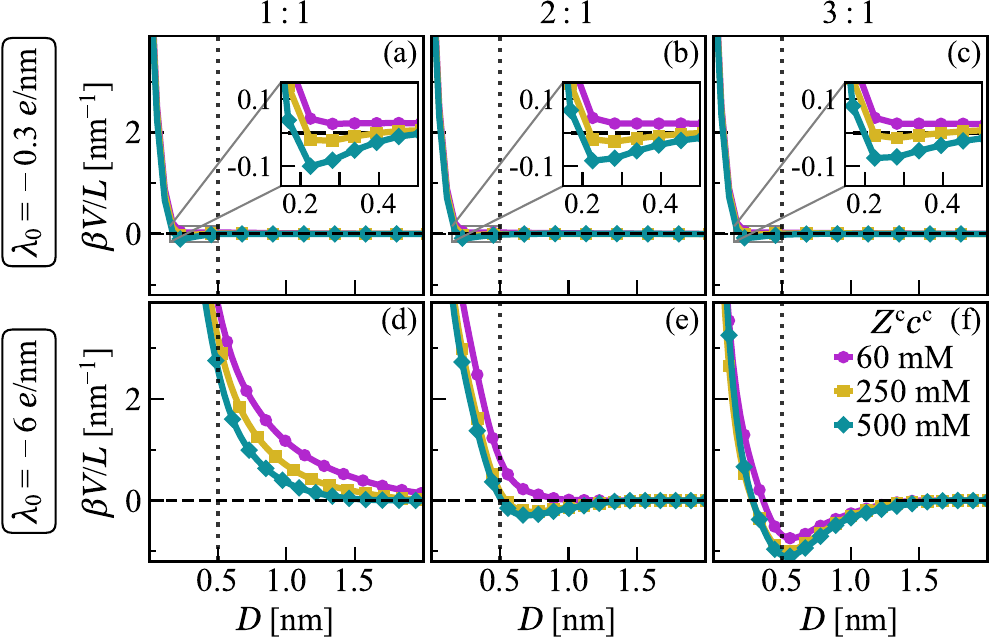}
 \caption{Normalized potential of mean force ($\beta V/L$) between PEs plotted against their surface-to-surface distance $D$. Data are presented for varying salt valency ratios ($Z^{\rm c}:\vert Z^{\rm a}\vert$, left to right), salt concentrations ($Z^{\rm c}c^{\rm c}$), and PE linear charge density of (a)-(c) $\lambda_0=-0.3$ $e/$nm and (d)-(f) $\lambda_0=-6$ $e/$nm. The substrate relative dielectric constant is $\varepsilon_{\rm s}=8000$. Both PE centers are fixed at $z=1.4$ nm from the substrate plane. Vertical dotted lines indicate $D=\sigma^{i}=0.5$ nm. }
 \label{two_PEs_PMF}
\end{figure}

In Fig.~\ref{two_PEs_PMF}, we present the normalized potential of mean force $\beta V/L$ as a function of the surface-to-surface distance $D$ between the PEs at $z=1.4$ nm.
Here, we systematically vary both the valency of cations in the added salt and its concentration.
We observe that the strongly charged PEs ($\lambda_0=-6$ $e/$nm) repel each other at all distances in $1:1$ salt solutions, and $V$ approaches zero at all concentrations when $D>2$ nm.
At high enough salt concentrations, when $Z^{\rm c}c^{\rm c}\geq 250$ mM, PEs attract each other in the presence of $2:1$ salt, see panel~(e).
This attractive force becomes even more pronounced in $3:1$ salt, where attraction appears at even lower concentrations (see panel~(f)).
This attraction is attributed to the bridging effect of the cations situated between the PEs and the correlation effects induced by these ions. 
Additionally, the results of Fig.~S5 of the SM show that altering the substrate dielectric constant $\varepsilon_{\rm s}$ from $500$ to $8000$ has no significant impact on the interactions between PEs near the high dielectric constant, metallic-character substrates, particularly at higher salt concentrations.

For weakly charged PEs ($\lambda_0=-0.3$ $e/$nm), we observe repulsion between PEs at low salt concentrations. However, at high salt concentrations, a subtle attraction is observed between nearly-touching PEs beyond the LJ interaction cutoff at $D>0.25$ nm.
The attraction rises from the entropic depletion effect of correlated ions~\cite{li2017, allahyarov1998} and is barely affected by the counterion valency, but increases with increasing salt (cf. Figs.~\ref{two_PEs_PMF}(a)-(c)). This is in contrast to the strongly charged PEs, where attraction comes from electrostatic interactions. 
In these cases, the depletion attraction rapidly approaches zero when $D>0.5$ nm.

\begin{figure}[b!]
\centering
  \includegraphics[width=1\linewidth]{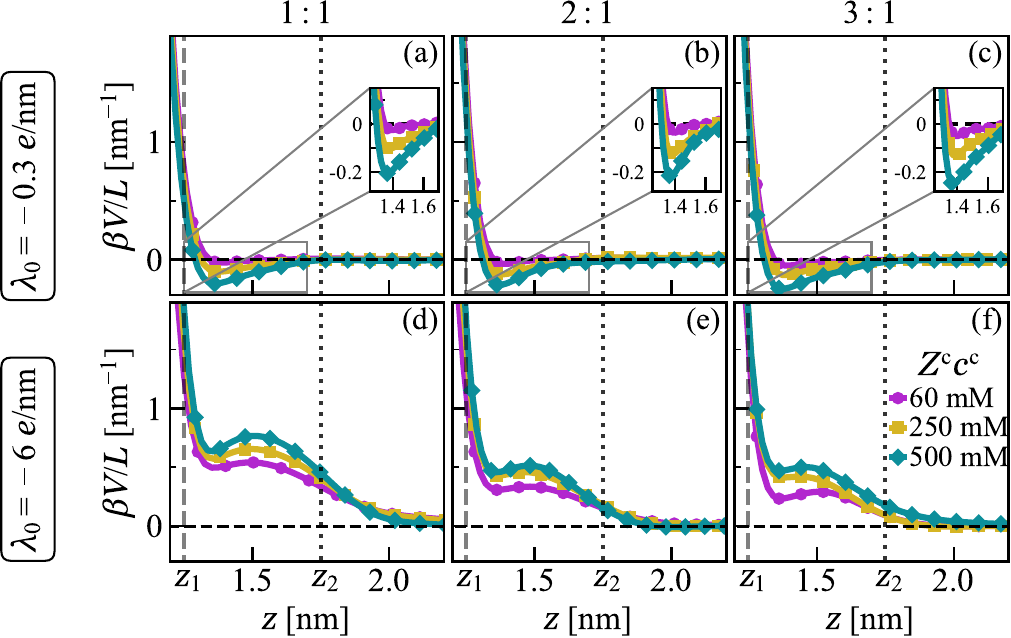}
 \caption{The normalized potential of mean force $\beta V/L$ as a function of the distance $z$ from the center of each PE to the substrate plane. The data are presented for salts with valencies $1:1$, $2:1$, and $3:1$ while varying the salt concentration $Z^{\rm c}c^{\rm c}$. The PE linear charge density is (a)-(c) $\lambda_0=-0.3$ $e/$nm and (d)-(f) $\lambda_0=-6$ $e/$nm.  
 The substrate relative dielectric constant is $\varepsilon_{\rm s}=8000$.
 The PEs are at their equilibrium distances of $D=0.25$ nm for $\lambda_0=-0.3$ $e/$nm and $D=0.55$ nm for $\lambda_0=-6$ $e/$nm from each other. Vertical dashed lines indicate where the PE surface contacts the substrate beads at $z_1=(\sigma^{\rm PE}+\sigma^{\rm s})/2$, and dotted lines show where the separation equals one ion diameter at $z_2=z_1+\sigma^i$.}
 \label{two_PEs_PMF_ZZ}
\end{figure}

\subsubsection{Interaction of a PE pair with a high dielectric constant substrate}

The relatively strong attraction induced by charge correlations due to multivalent counterions leads to bridged PE chains together in the bulk of the solution.
On the other hand, strongly charged PEs experience screening-reduced adsorption with increasing salt concentration, as shown in Fig.~\ref{lambda_PE}.
The remaining question then concerns the influence of the second PE on the PE-substrate interactions.
To obtain a dense adsorbate layer, an attractive interaction towards the substrate and a potential energy minimum should prevail upon adsorption on the substrate. To this end, we consider two PEs approaching the substrate at their corresponding minimum energy distances from each other at $D=0.25$ nm and $D=0.55$ nm for PE charge densities of $\lambda_0=-0.3$ $e/$nm and $\lambda_0=-6$ $e/$nm, respectively. 
In cases where PEs do not exhibit mutual attraction, such as for $\lambda_0=-6$ $e/$nm in $1:1$ salt, we use the same $D$ value used for other salts at that particular $\lambda_0$.

Figure~\ref{two_PEs_PMF_ZZ} shows that the strongly charged PEs ($\lambda=-6$ $e/$nm) are repelled from the substrate at all salt concentrations here.
The competition between electrostatic attraction and steric repulsion governs the adsorption process here. 
Initially, condensation of cations between the two PEs (see Fig.~\ref{Fig2-schematic}(c)-(e)) results in the neutralization of charges induced on the substrate, thereby diminishing the electrostatic attraction towards the substrate.
Increasing the cation valency lowers the steric barrier height at every salt concentration, possibly due to the reduced number of ions in the system.
In addition, the steric repulsion, arising from the condensed ions between the substrate and the PEs, becomes the dominant force, leading to repulsion of the PEs.
This aligns with theoretical findings using PB theory, which have shown that DNA-cationic lipid complexes with monovalent counterions become unstable when additional DNA strands are adsorbed onto the substrate~\cite{bruinsma1998, cherstvy2007}.
We note that a higher salt concentration enhances the ion condensation on the PEs, thereby augmenting this repulsive effect.   
On the other hand, PEs with a low charge density $\lambda=-0.3$ $e/$nm can be adsorbed to the substrate even for monovalent salt, and the adsorption strength increases as the salt concentration increases.
In this case, similar to the single PE case shown in Fig.~\ref{lambda_PE}, there is no energy barrier impeding the adsorption of the PEs to the substrate due to the reduced attraction of cations to the PEs.   
Furthermore, the entropic depletion effect leads to the PEs being attracted to the substrate, an effect that enhances with the rise in salt concentration.
Similar findings were reported in Ref.~\onlinecite{harmat2023}, where interactions of poly-L-lysine and poly-L-arginine with $\alpha$-quartz surfaces were studied, with sodium ions as counterions. It was observed that at low pH levels, where the peptides are strongly charged, significant electrostatic repulsion between peptides leads to reduced substrate coverage. Conversely, at higher pH levels, the reduced charge on the peptides enhances the formation of denser peptide films. Despite increased substrate coverage, these films exhibit decreased stability.
Our results in Fig.~\ref{two_PEs_PMF_ZZ} further show that increasing cation valency to $Z^{\rm c}=3$ does not notably affect the interaction between the PEs and the substrate.

\begin{figure}[b!]
\centering
  \includegraphics[width=1\linewidth]{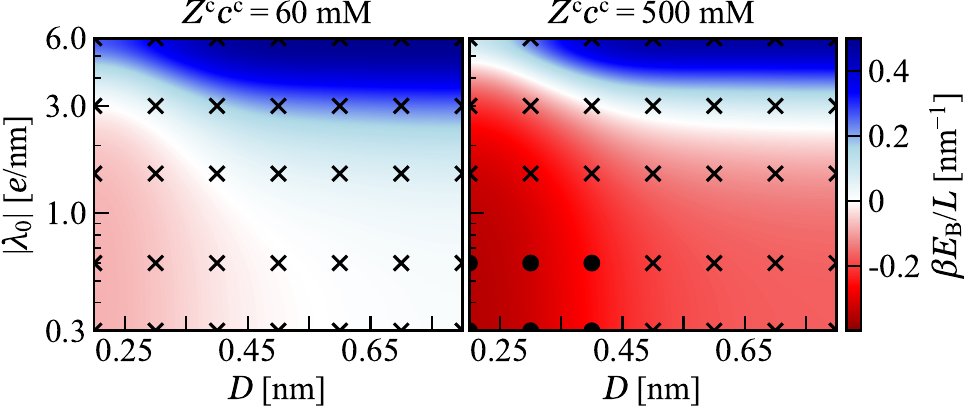}
 \caption{The normalized binding energy of the PEs to the substrate $\beta E_{\rm B}/L$ as a function of PE linear charge density $\lambda_0$ and surface-to-surface PE distance $D$. The data correspond to $1:1$ salt at concentration $Z^{\rm c}c^{\rm c}=60$ mM (left) and $Z^{\rm c}c^{\rm c}=500$ mM (right).
 The substrate relative dielectric constant is $\varepsilon_{\rm s}=8000$. Simulated values are marked with symbols; circles represent cases where the PEs attract each other, and crosses indicate cases where the PEs repel each other. The color scheme is based on 2D interpolation between the discrete measurements. The maximum value of the color bar is capped at $0.6$ nm$^{-1}$. The higher values are shown in dark blue. Figure~S8 of the SM presents $\beta V/L$ between PEs and the substrate as a function of their distance $z$ for selected points from this figure.}
 \label{interpolation}
\end{figure}

To summarize the findings of this section, we calculate the binding energy $E_{\rm B}$, which is defined as the energy required to bring a bound PE pair from their equilibrium distance close to the substrate to infinity and can be quantified by twice the value of the (negative) minimum of $V$ in
Fig.~\ref{two_PEs_PMF_ZZ}, where $V_{\rm min} = V(z_{\rm min})$, with $z_{\rm min} \approx 1.4$ nm.
This distance is the equilibrium distance for cases where adsorption to the substrate is observed. 
For consistency, the same value of $z_{\rm min}$ is also used for cases where adsorption is not observed.
In Fig.~\ref{interpolation}, we explore the dependence of $E_{\rm B}$ on both $D$ and $\lambda_0$ in $1:1$ salt solution at two concentrations in the case of a high dielectric constant, metallic-character substrate with $\varepsilon_{\rm s}=8000$. This case has been chosen since, from the data presented above, it is the most promising setup for the production of a densely populated monolayer.
At the lowest salt concentration $Z^{\rm c} c^{\rm c}=60$ mM (left), the PE-substrate interaction is attractive when $\vert \lambda_{0}\vert<3$ $e/$nm, and the PEs are sufficiently close, i.e., $D<0.45$ nm. However, the PEs themselves do not attract each other (cf. Fig.~\ref{two_PEs_PMF}(a)).
At the highest salt concentration $Z^{\rm c} c^{\rm c}=500$ mM (right), the PEs are attracted to the substrate when $\vert \lambda_{0}\vert<1.5$ $e/$nm for all $D$. On the other hand, PE-PE attraction is observed when $D<0.45$ nm and $\vert \lambda_{0}\vert<0.6$ $e/$nm (cf. Fig.~\ref{two_PEs_PMF}(a) and Fig.~S7 of the SM), denoted with circles in Fig.~\ref{interpolation}. These relatively few points are the only cases for which the formation of a dense monolayer can be expected on a high dielectric constant, metallic-character neutral substrate.

\subsection{Single PE perpendicular to the substrate}

The electrophoretic translocation of DNA and other biomolecules through membrane nanopores is of significant importance in biotechnology and has recently attracted attention~\cite{palyulin2014, buyukdagli2019}.
A key aspect of sequence reading during translocation is the regulation of DNA movement~\cite{clarke2009}, which necessitates a quantitative characterization of DNA-medium interactions.
Here, we investigate a DNA-like rod molecule oriented perpendicular to the substrate, while varying the dielectric permittivity $\varepsilon_{\rm s}$ and salt concentration. This model is relevant for studying the DNA approach stage towards the membrane when controlled by purely electrostatic forces. The PE here has a length of $10$ nm with $\lambda_0 = -6$ {\it e}/nm. For this case, we vary the substrate dielectric constant as in the Sec.~\ref{sec:results_discussion} of this manuscript. 
The PE approach stage towards a thin membrane has been analytically studied using field-theoretical approaches beyond the MF theory to highlight the influence of dielectric mismatch on polymer translocation through membranes~\cite{buyukdagli2016, buyukdagli2016-2, buyukdagli2018}.
It was shown that polymers passing through carbon-based membranes with $\varepsilon_{\rm s}=2$ face a significant Coulombic repulsion. Conversely, membranes engineered to have a high dielectric constant are strongly attractive and can trap the PE in the pore. In our system, we study the approach on the {\it cis} side of the substrate only because the substrate medium virtually fills the whole region for $z < 0$.
\begin{figure}[t!]
\centering
\includegraphics[width=1\linewidth]{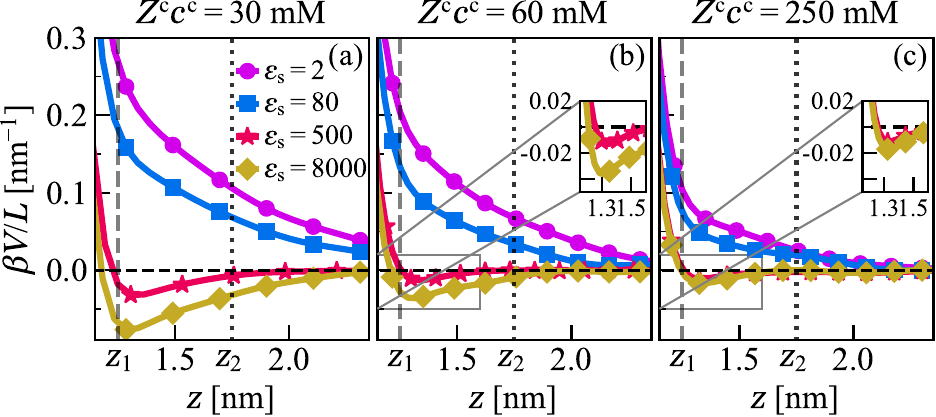}
 \caption{The normalized potential of mean force $\beta V/L$ as a function of the distance $z$ between the substrate and the center of the PE bead closest to it in $1:1$ salt with the concentration of (a) $Z^{\rm c}c^{\rm c}=30$ mM, (b) $Z^{\rm c}c^{\rm c}=60$ mM, and (c) $Z^{\rm c}c^{\rm c}=250$ mM for varying relative dielectric constant of the substrate $\varepsilon_{\rm s}$. The vertical dashed lines represent the distance at which the PE head bead contacts the substrate surface at $z_1=(\sigma^{\rm PE}+\sigma^{\rm s})/2$, while the dotted lines represent the distance at which they are one ion diameter apart at $z_2=z_1+\sigma^i$.}
 \label{one_PE_perpendicular}
\end{figure}

In Fig.~\ref{one_PE_perpendicular}, we present $\beta V/L$ between the substrate and the DNA as a function of the distance between the substrate plane and the center of the PE bead closest to it.
As expected, $\beta V/L$ in this configuration is significantly smaller than the values observed for PEs oriented parallel to the substrate.
This observation also suggests that without the perpendicular constraint, the parallel orientation of the PE is energetically more favorable than the perpendicular one when the interaction is attractive.
Similarly to the case where PEs are aligned parallel to the substrate, substrates with a dielectric constant $\varepsilon_{\rm s}$ smaller than that of water $\varepsilon_{\rm w}$ repel the PE, whereas substrates with a $\varepsilon_{\rm s}$ higher than $\varepsilon_{\rm w}$ attract the PE. 
In general, increasing the salt concentration weakens the PE-substrate interactions as a result of screening.

Monte Carlo simulations have demonstrated that rodlike PEs in the vicinity of an oppositely-charged substrate with $\Delta \varepsilon=0$ tend to align parallel to the substrate~\cite{messina2006}.
Our results support the observed behavior close to substrates with $\varepsilon_{\rm s}=8000$, where the attraction of the PE to the substrate is stronger when the PE is aligned parallel to the substrate as compared to the perpendicular orientation (cf. Figs.~\ref{Fig1} and~\ref{one_PE_perpendicular}).  
Moreover, theoretical investigations revealed that perpendicular orientation of a PE near a like-charged substrate with $\Delta\varepsilon=0$ is more favorable than parallel orientation~\cite{buyukdagli2019-2, sahin2019}, and this preference is further enhanced by increasing counterion valency~\cite{buyukdagli2019-2}. Consistently, Fig.~\ref{one_PE_perpendicular} shows that the repulsive force on the PE near a substrate with $\varepsilon_{\rm s}=2$ is weaker as compared to the repulsion observed for a parallel orientation (see Fig.~\ref{Fig1}). This indicates a shift in the orientation of the PE, from being perpendicular to becoming parallel to the substrate near its surface.

\section{Summary and Conclusions}\label{conclusion}

In this work, we have used a simple coarse-grained MD model to assess the spontaneous adsorption of rigid rodlike PEs on uncharged substrates in implicit water. We demonstrated that the adsorption of PEs to the substrates can be regulated via the substrate dielectric constant, salt concentration, PE charge, and ion valency. As expected,
when the substrate dielectric constant $\varepsilon_{\rm s}$ is lower than that of water $\varepsilon_{\rm w}$, PEs are repelled from the substrate due to the presence of image charges with the same sign.
On the contrary, when the substrate has a high dielectric constant as compared to that of water ($\varepsilon_{\rm s}>\varepsilon_{\rm w}$), the induced image charges have opposite sign to the PEs leading to attraction between the PEs and the substrate which may facilitate adsorption.

As added salt concentration increases, the resulting behavior becomes more complex due to the interplay of steric and electrostatic forces between the substrate and the PE~\cite{drechsler2010}.
If the PE is strongly charged, the condensed counterion layer in the region between the PE and the substrate swells, leading to significant steric repulsion~\cite{netz1999}.
For weakly charged PEs, the electrostatic interactions of the PE with counterions are weak, and the condensed counterion layer collapses and reduces the steric barrier between the PE and the substrate. Consequently, at the smallest distances between the substrate and the PE, the interactions are mainly governed by electrostatic and entropic depletion forces that attract the PEs to the substrate~\cite{alarcon2013}, especially at high salt concentrations.

To explore the possibility of forming dense monolayers of like-charged PEs on metallic-character
substrates, systems with two PEs were investigated. Here, the steric barrier against the adsorption of strongly charged PEs was pronounced, leading to their repulsion from the substrate across different salt valences and concentrations. In contrast, when PEs are weakly charged, depletion-driven attraction becomes more dominant than steric repulsion. Pairs of weakly charged PEs exhibited an attraction to the substrate, similar to the behavior of a single PE, at sufficiently high salt concentrations ($Z^{\rm c}c^{\rm c}>250$ mM). Moreover, at these concentrations, PEs attract each other.
Figure~\ref{interpolation} provides a summary of our results, illustrating the specific ranges of $\lambda_0$ and PE-PE distances where attraction (in red) and repulsion (in blue) between the PEs and the substrate are observed. The setup chosen for this exploration corresponds to the most promising setup for the formation of a monolayer from the combined data obtained in Figs.~\ref{two_PEs_PMF} and \ref{two_PEs_PMF_ZZ}.
Our main conclusion is that cases for which the formation of a dense monolayer can be expected on a high dielectric constant, metallic-character substrate are few, marked as circles in Fig.~\ref{interpolation}. This provides guidelines for the delicate process of system engineering aiming at like-charged PEs forming closely packed films on neutral, polarizable substrates.
Finally, we note that the PE length must be sufficiently long such that the binding energy $E_{\rm B} \gg k_{\rm B}T$ to prevent desorption due to thermal fluctuations.

Finally, we have investigated the interactions of a PE perpendicular to the substrate. Our results reveal that the potential of mean force in this geometry is significantly smaller than in the case where the PE is parallel to the substrate, hinting at a preference for parallel orientation in free-chain configurations.

Overall, our comprehensive analysis provides valuable insights into the complex interplay of electrostatic forces, ion distribution, and substrate properties in determining the behavior of PEs near substrates. These findings not only enhance our understanding of PE-substrate interactions but also have implications for a range of applications, from biotechnology to material science. It would also be interesting to extend our approach to flexible chains on charged substrates.

Our work has focused on the interaction of rigid PEs with planar flat substrates.
In this geometry, image charge calculation results in effective surface element charges unaffected by the induced charges on neighboring elements due to the condition $\vect{n}\perp \vect{E}^i$. 
In future works, it would be interesting to consider other substrate geometries, such as convex or concave substrates interacting with flexible PEs.
In such cases, the image charge calculation is influenced by the system geometry and the charge distribution~\cite{wu2018}.
This allows, e.g., curvature controlled creation of diverse surface charge patterns~\cite{pogharian2024}.
For example, in salt solutions in contact with a substrate with $\varepsilon_{\rm s}<\varepsilon_{\rm w}$, the repulsive forces on ions weaken near convex regions and strengthen near concave ones~\cite{wu2018, solis2021}.
Also, a geometry change from an insulating planar substrate to a spherical shape results in a decrease of image-charge repulsion between a flexible PE and the substrate surface~\cite{cherstvy2012}.

Our modelling results can aid in the interpretation of experimental data and enhance the performance of applications.
For example, PE elastomers have widespread applications in flexible electronic devices, such as polymer-based transducers~\cite{pavlenko2016} and soft robotics~\cite{nguyen2024}, but also in energy harvesting~\cite{guo2015}, energy storage~\cite{qiu2022}, and in printable electronics~\cite{tong2022}.
In these applications, the dielectric mismatch between the substrate and the PE solutions affects device efficiency and performance.
For example, metal oxide electrodes with higher dielectric constant significantly improve the capacitance and the electrochemical efficacy of supercapacitors~\cite{tu2020}. 
An interesting application direction for materials with high dielectric constants is ingestible electronics and edible composites made from natural biopolymers~\cite{konwar2022}.
Characterization of PE-substrate interactions is also important for optimizing other applications, including biosensors~\cite{hensel2020}, artificial cells~\cite{xu2016}, drug delivery systems~\cite{zhang2018}, and self-healing coatings~\cite{abu2016}.

\section*{Acknowledgments}
This work was supported by Finnish Cultural Foundation under grant no. 00241182 (H.V.). The work was also supported by the Academy of Finland through its Centres of Excellence Programme (2022-2029, LIBER) under project nos. 346111  and 364205 (M.S.), Academy of Finland project no. 359180 (M.S.) and under the European Union – NextGenerationEU instrument by the Academy of Finland grant 353298 (T.A-N.). The work was further supported by Technology Industries of Finland Centennial Foundation TT2020 grant (T.A-N.).
We are grateful for the support by FinnCERES Materials Bioeconomy Ecosystem. Computational resources by CSC IT Centre for Finland, RAMI -- RawMatters Finland Infrastructure and the Aalto Science-IT are also gratefully acknowledged.

\section*{REFERENCES}
\bibliographystyle{aipnum4-1}

\bibliography{references}
\end{document}